# Quasi-Fermi level splitting in nanoscale junctions from *ab initio*

Juho Lee, Hyeonwoo Yeo, and Yong-Hoon Kim*

School of Electrical Engineering and Graduate School of EEWS, Korea Advanced Institute of Science and Technology (KAIST), Daejeon, 305-701, Korea

*Corresponding author: y.h.kim@kaist.ac.kr

**ABSTRACT:** The splitting of quasi-Fermi levels (QFLs) represents a key concept utilized to describe finite-bias operations of semiconductor devices, but its atomic-scale characterization remains a significant challenge. Herein, the non-equilibrium QFL or electrochemical potential profiles within single-molecule junctions obtained from the newly developed first-principles multi-space constrained-search density functional formalism are presented. Benchmarking the standard non-equilibrium Green's function calculation results, it is first established that algorithmically the notion of separate electrode-originated nonlocal QFLs should be maintained within the channel region during self-consistent finite-bias electronic structure calculations. For the insulating hexandithiolate junction, the QFL profiles exhibit discontinuities at the left and right electrode interfaces and across the molecule the accompanying electrostatic potential drops linearly and Landauer residual-resistivity dipoles are uniformly distributed. For the conducting hexatrienedithiolate junction, on the other hand, the electrode QFLs penetrate into the channel region and produce split QFLs. With the highest occupied molecular orbital entering the bias window and becoming a good transport channel, the split QFLs are also accompanied by the nonlinear electrostatic potential drop and asymmetric Landauer residual-resistivity dipole formation. Our findings underscore the importance of the first-principles extraction of QFLs in nanoscale junctions and point to a new direction for the computational design of next-generation electronic, optoelectronic, and electrochemical devices.

## 1. INTRODUCTION

The quasi-Fermi (imref) level (QFL) concept, which was originally proposed by Shockley,[1-3] is a standard construct for describing finite-bias non-equilibrium operations of semiconductor electronic, optoelectronic, and electrochemical devices.[4-5] Because of the direct relationship with the non-equilibrium charge transport processes in nanoscale junctions,[6-7] there have been also intensive on-going efforts to explicitly measure the spatial and energetic distributions of QFLs in the form of the local electrochemical potential at the atomic scale.[8-10] In clarifying the correlations between the QFL profiles and localized scatterers like interfaces, steps, and dopants within current-flowing nanostructures and further designing advanced nanodevices, first-principles non-equilibrium quantum transport calculations can potentially play an important role. However, the standard approach combining density-functional theory (DFT) with non-equilibrium Green's function (NEGF) formalism within the Landauer picture has been less successful in extracting the QFL profiles because it does not explicitly provide information on single-electron wavefunctions and their occupations.[11-12]

Herein, carrying out first-principles non-equilibrium electronic structure and quantum transport calculations within the multi-space constrained-search DFT (MS-DFT) formalism we have recently developed,[13] we report the first-principles calculations of QFL profiles in single-molecule junctions and their correlations with finite-bias transmissions. For nanoscale constrictions where the Landauer picture can be invoked, the MS-DFT method is established by (i) adopting a micro-canonical picture that considers finite electrodes (rather than semi-infinite electrodes within the Landauer picture) and (ii) mapping the finite-bias quantum transport process to the drain-to-source multi-electrode electronic excitation (optical analogy). A key step of MS-DFT is to variationally determine the occupations of the spatially-resolved single-electron Kohn-Sham eigenstates within the constraint of $eV_b = \mu_L - \mu_R$, and accordingly the QFLs are explicitly and straightforwardly produced within the approach (see also **Materials and Methods**). We first carry out DFT-based NEGF (DFT-NEGF) calculations for single-molecule junctions based on Au-S linkages[14] and prepare the benchmark data. As molecules, we consider, in addition to the experimentally and theoretically well-studied saturated hexane-1,6-dithiolate (S-$C_6H_{12}$-S) molecule,[15] the conjugated 1,3,5-hexatriene-1,6-dithiolate (S-$C_6H_6$-S) molecule that has important implications for bioprocesses.[16] Comparisons with the DFT-NEGF calculation results then enables us to determine the occupation rule of the spatially-resolved Kohn-Sham states or the QFLs within the MS-DFT approach. Next, analyzing the QFL profiles of the insulating hexanedithiolate (HDT) junction, we observe the left and right electrode electrochemical potentials abruptly terminating at the respective metal-molecule interfaces and the corresponding electrostatic potential linearly dropping across the molecule. On the other hand, for the conducting hexatrienedithiolate (HTDT) case, we find that the electrochemical potentials of the two electrodes extend into the molecular region and together with the highest occupied molecular orbital (HOMO) state of the HTDT form energetically split QFLs. Becoming a dominating charge transport channel within the bias window, the contribution of the HTDT HOMO to the electrochemical potentials is accompanied by an asymmetric non-linear electrostatic potential drop across the molecule. We thus clarify the nontrivial correlations between the electrochemical and electrostatic potential drop profiles at the atomic scale.

## 2. RESULTS AND DISCUSSION

**Electron occupation rule for non-equilibrium channel states.** While the electrochemical potentials or QFLs are in principle defined as the occupation of non-equilibrium energy levels, it is in practice a non-trivial matter to compute them in a first-principles manner. It should be noted that in the standard DFT-NEGF formalism,[6-7] the explicit determination of QFLs is avoided as follows: First, DFT-NEGF invokes the Landauer picture in which one considers an open quantum system composed of a channel $C$ sandwiched by the left electrode $L$ and the right electrode $R$, as illustrated in Fig. 1A. The left and right electrodes are in local equilibrium but maintained at two different electrochemical potentials $\mu_L$ and $\mu_R$, respectively, by the applied external bias, $V_b = (\mu_L - \mu_R)/e$. Then, for the channel driven into a non-equilibrium state, the electron correlation function $G^n$ (or lesser Green's function $G^<$) that contains the information on the density matrix is calculated as a function of energy $E$ according to

$$G^n(E) \equiv -iG^<(E) \approx A_L(E)f_L(E) + A_R(E)f_R(E), \quad (1)$$

where $A_{L/R}$ is the channel spectral function portion associated with the electrode $L/R$, and $f_{L/R}$ is the Fermi function associated with the electrode $L/R$,

$$f_{L/R}(E) \equiv f(E - \mu_{L/R}) = \frac{1}{1+\exp[(E-\mu_{L/R})/k_BT]}. \quad (2)$$

Note that, because the electron states are not explicitly generated within DFT-NEGF, QFLs have been extracted as a post-processing procedure based on certain approximations on the form of the averaged local electrochemical potential.[11-12, 17]

Within the MS-DFT formalism, we now adopt finite-sized $L$ and $R$ electrodes (micro-canonical ensembles) and explicitly extract the Kohn-Sham (KS) states by performing the multi-space constrained-search procedure or total-energy minimizations with the constraint of $eV_b = \mu_L - \mu_R$. A key step is then, during the self-consistent construction of non-equilibrium charge densities, variationally occupying the KS states with a proper occupation rule. In establishing the occupation rule, a reasonable approach would be to assume that the eigenstates of the $L$-$C$-$R$ system can be divided into one group originating from $L$ and right-traveling with a population function $f_L$ and the other group originating from $R$ and left-traveling with a population function $f_R$. Note that this more or less corresponds to the scheme that was implicitly used in the implementation of DFT-NEGF in the form of equation [1]. To embody this rule within MS-DFT, we thus for each $C$ state (i.e. the state with the biggest weight in region $C$) assessed the $L$ and $R$ weights and filled the level according to (Fig. 1B top panel)

Occupation rule **I**:

$$f_C \to \begin{cases} L - \text{originated:} & f(E - \mu_L) \\ R - \text{originated:} & f(E - \mu_R) \end{cases}. \quad (3)$$

Rather than dealing with multiple split QFLs explicitly, one might also want to consider only a single averaged electrochemical potential. So, we additionally tested the equal-weight averaging occupation rule (Fig. 1B bottom panel),

Occupation rule **II**:

$$f_C \to f(E - \frac{\mu_L + \mu_R}{2}), \quad (4)$$

which could be a representative scheme of generating the local electrochemical potential that was found to be useful for post-processing analysis.[11-12, 17]

Carrying out the MS-DFT calculations for the single-molecule junction models based on the HDT and HTDT molecules and benchmarking the DFT-NEGF calculation results, we now heuristically determine the validity of the above two occupation rules. In modeling the junctions (Fig. 1C), we considered the S-Au linkages based on a single apex Au atom or coordination number one S-Au contacts, for which the molecular states are energetically pulled up as close as possible to the equilibrium Fermi level $E_F$.[18] In carrying out the self-consistent DFT-NEGF and MS-DFT calculations, we applied the Fermi functions with a temperature of 300 K to both $L$ and $R$ reservoirs. In Fig. 1D, we show for the HDT and HTDT junctions under $V_b = 1.0$ V and 0.6 V, respectively, the plane-averaged bias-induced electrostatic potential change profile,

$$\Delta \bar{v}_H(\vec{r}) = \bar{v}_H^V(\vec{r}) - \bar{v}_H^0(\vec{r}), \quad (5)$$

where $\bar{v}_H^V$ and $\bar{v}_H^0$ are the classical Hartree Coulomb potentials at the non-equilibrium and equilibrium conditions, respectively. The corresponding exchange-correlation potential variations are much smaller and will not be explicitly discussed below (see *SI Appendix*, Fig. S1). It can be then immediately found that, for both the HDT and HTDT junction cases, the DFT-NEGF $\Delta \bar{v}_H$ curves (black dotted lines) are accurately reproduced by the MS-DFT $\Delta \bar{v}_H$ ones only within the occupation rule **I** that explicitly maintains two separate $\mu_L$ and $\mu_R$ (red solid lines). On the other hand, with the averaging occupation rule **II**, we obtain the MS-DFT $\Delta \bar{v}_H$ (blue solid lines) that significantly deviate from the DFT-NEGF $\Delta \bar{v}_H$. It should be commented here that, for the HDT junction case, the adaptation of the coordination number three S-Au contact or shifting the HOMO further away from $E_F$ and out of the bias window[18] enabled the DFT-NEGF $\Delta \bar{v}_H$ to be reproduced by MS-DFT within both the occupation rules **I** and **II** (see *SI Appendix*, Fig. S2). As will be discussed shortly, the HDT junction behaves as a simple tunneling barrier, and the HOMO originates from the spatially localized S states. We thus conclude that the average occupation approach **II** is appropriate only for insulating junctions with simple electrode-channel interfacial electronic structures. We consider that these findings are in line with the earlier claim on the validity of the local electrochemical potential concept only in the linear-response regime or low-temperature and low-bias conditions.[17] Having confirmed that algorithmically it is essential to maintain the notion of two separate $\mu_L$ and $\mu_R$ during self-consistent non-equilibrium electronic structure calculations, we will below present only the MS-DFT calculation results obtained with the occupation rule **I**.

**Non-equilibrium quantum transport characteristics.** Next, we examined in detail the quantum transport properties of the two molecular junctions. In Figs. 2A and 2B, we first present the current-bias voltage ($I$-$V_b$) characteristics of HDT and HTDT junctions, respectively, which show the insulating character of the former and the conducting character of the latter. While alkane-dihiolate molecular junctions have been extensively investigated both experimentally and theoretically,[14-15] studies on polyene-dithiolate counterparts are scarce[16, 19] despite the important roles played by polyenes for bioprocesses such as vision and photosynthesis.[20] Due to the conjugated nature of HTDT or the presence of three double bonds, the HTDT junction produces currents about two orders of magnitude larger than those from its HDT counterpart.

To understand this large discrepancy between their current-carrying capacities, we observe the transmission functions and the corresponding molecule-projected density of states (DOS) obtained at $V_b = 0.6$ V. The very different $I$-$V_b$ characteristics are the direct results of negligible transmissions in the HDT junction within the bias window (Fig. 2C left panel) except for single near-unity transmission peak right below the middle of the bias window $(\mu_L + \mu_R)/2$ in the HTDT case (Fig. 2D left panel). However, the projected DOS data show that the HOMO enters the bias window in the HDT junction (Fig. 2C right panel, left filled triangle) as in its HTDT counterpart (Fig. 2D right panel, left filled triangle).

These seemingly contradictory results can be understood in terms of the spatial distributions of their HOMOs, and for this purpose, we present in Figs. 2E and 2F the molecular-projected Hamiltonian (MPH) analysis[21] data and in Figs. 3A and 3B the spatially-resolved DOS. The local DOS of the HDT junction (Fig. 3A) then show that the equilibrium HDT HOMOs are initially composed of two states localized at the left and right S sites, and at finite $V_b$, due to their disconnected nature, the left and right interfacial states are independently pulled up and down, respectively. The localized character of the left S-based HOMO, which corresponds to the DOS peak within the bias window in Fig. 2C right panel (indicated by the left triangle), can be also clearly observed in the MPH state visualization shown in Fig. 2E.

Similarly, we can establish the correlation between the spatial distribution of HOMO and the conducting nature of the HTDT junction: The most notable feature from the MPH visualization shown in Fig. 2F and the local DOS plot shown in Fig. 3B is the highly delocalized nature of the HTDT HOMO-originated state (left triangle in Fig. 2D right panel). Due to this delocalization, in contrast to the splitting of the HOMO into two decoupled states in the HDT case (Figs. 2C and 3A), the equilibrium HOMO maintains its identity even at finite $V_b$ (single DOS peak in Fig. 2D right panel; see also Fig. 3B) and produces a near-unity transmission peak for the HTDT junction (Fig. 2D left panel).

**Equivalence and comparison of DFT-NEGF and MS-DFT.** At this point, we emphasize that, as shown in Fig. 2,

the $I$-$V_b$, transmission, DOS, and MPH visualization data obtained based on MS-DFT are in excellent agreement with the DFT-NEGF data in both the insulating HDT and conducting HTDT junction cases (see also *SI Appendix,* Fig. S2). While we previously reported the accordance between DFT-NEGF and MS-DFT calculation results for van der Waals heterostructures,[13] the demonstration of such correspondence here for covalently bonded molecular junctions reinforces the validity of adopting the MS-DFT formalism to describe non-equilibrium quantum transport processes.

Before presenting the electrochemical potentials uniquely available in MS-DFT, we discuss several computational outputs that can be commonly obtained from the DFT-NEGF and MS-DFT calculations. First, we point out that the MPH eigenstates shown in Figs. 2E and 2F were obtained by diagonalizing a non-equilibrium Hamiltonian matrix derived from a cluster system, namely by introducing an arbitrarily defined supercell that contains a vacuum and with a Γ-point sampling, as a post-processing step. We will present below the KS states obtained from self-consistent MS-DFT calculations, which more clearly exhibit the bulk crystal wavefunction characteristics in the electrode regions.

Next, we consider the bias-induced electrostatic potential changes and the corresponding charge density differences. Comparing the $\Delta \bar{v}_H$ profiles of the HDT and HTDT junctions shown as black solid lines in Figs. 3A and 3B as well as Fig. 1D, we note that, while it drops uniformly across the molecule in the former (Fig. 1D top panel and Fig. 3A), $\Delta \bar{v}_H$ drops more abruptly on the left electrode-molecular interface in the latter (Fig. 1D bottom panel and Fig. 3B). The sharp electrostatic potential drop at the left S site for the HTDT junction case is more noticeable in the contour plot representations shown in the bottom panels of Figs. 3C and 3D for the HDT and HTDT junctions at $V_b$ = 0.6 V, respectively. To understand their implications, we additionally analyzed the bias-induced charge density difference,

$$\Delta \rho(\vec{r}) = \rho^V(\vec{r}) - \rho^0(\vec{r}), \quad (6)$$

where $\rho^V$ and $\rho^0$ are the charge densities of each junction at the non-equilibrium and equilibrium conditions, respectively. The contour plots of $\Delta \rho$ of the HDT and HTDT junctions at $V_b$ = 0.6 V are shown in the top panels of Figs. 3C and 3D, respectively, and we find that $\Delta \rho$ is rather uniformly distributed across the molecule in the HDT junction case but is concentrated near the left S site in the HTDT counterpart. Corresponding to the Landauer residual-resistivity dipole,[22-23] they then translate into the linear voltage drop across the channel in the HDT junction and the abrupt voltage drop at the left electrode-channel contact in the HTDT junction.

Here, it should be reminded that, in addition to the electrostatic potential $\Delta v_H$ drop, the question of where the "voltage drop" can be also addressed in terms of the electrochemical potential $\mu$ drop.[17, 23] Their spatial variations can deviate from each other within the non-equilibrium nanoscale channel region, and the discrepancies between electrostatic and electrochemical potential drops will become clearer in the subsequent discussions of the QFLs that are uniquely available within the MS-DFT approach.

**First-principles determination of QFLs.** While we established the correlations between electrostatic potential drops and Landauer residual-resistivity dipole distributions at finite bias voltages, the microscopic origin of the formation of a strong resistivity dipole on the left S site in the HTDT junction case remains unclear. We now demonstrate that the consideration of QFL profiles, or the individual states within the bias window and their occupations, unambiguously explains the asymmetric charge pileup behavior despite the symmetric contact atomic geometries. We show the MS-DFT-derived QFL profiles of the HDT and HTDT junctions under $V_b$ = 0.6 V in Figs. 4A and 4B, respectively. Here, we chose to represent the QFL profiles in terms of the xy-plane-projected KS wavefunctions together with the corresponding color-coded electron occupation factors obtained within the MS-DFT formalism. In selecting the KS states to be visualized, we extracted the Γ–point states with the electron occupancy values between 0.001 and 0.999.

If we first consider the HDT junction case (Fig. 4A), reflecting the simple tunneling barrier nature of HDT (negligible transmissions within the bias window in Fig. 2C left panel), QFLs are absent within the molecular backbone region. Analyzing the QFL states, we observe both states localized at the Au surface regions (indicated by grey arrows) and those delocalized throughout the Au region and connected to the S states (for the details, see *SI Appendix,* Fig. S3). We find that, while the former localized states support the linear electrostatic potential drop (black solid line), the latter states propagate from the deep $L$ and $R$ Au regions into the left and right S sites at $\mu_L$ (marked as ①) and $\mu_R$ (marked as ②), respectively. Visualization of the ① and ② states in Fig. 4C more intuitively reveal the spatially discontinuous nature of the QFLs.

Moving to the HTDT junction QFL profiles (Fig. 4B), we find that much richer and significantly different features emerge. Within the Au electrode regions, compared with the HDT junction case, more left (right) electrode states near the $\mu_L$ ($\mu_R$) energy range participate in forming QFLs. More importantly, within the channel region, we observe that the QFLs split into the HTDT HOMO (marked as ② in Fig. 4B) as well as the left- and right-electrode-originated states (① and ③ in Fig. 4B, respectively; for the details, see *SI Appendix,* Fig. S4). While the QFL splitting was introduced in the early development of semiconductor device physics for the basic concepts such as the *p-n* junction,[1] electron-hole recombination,[2] and solar cell,[3] to our knowledge, its first-principle determination has not been reported yet.

Based on the spatial distributions of the three states ①, ②, and ③ and their energetic locations (Figs. 4B and 4D), we finally explain the atomistic origins of the near-unity transmission peak (Fig. 2D left panel) and the nonlinear electrostatic potential drop in the HTDT junction (solid black line in Fig. 4B). First, we can see that the delocalization of the HTDT HOMO ② into both the right and left Au electrode regions results in an almost perfect transmission (Fig. 4D middle panel). In addition, because the HTDT HOMO energetically lies more closely to $\mu_R$, we observe that the coupling of the HTDT HOMO with the *R* states becomes much stronger than that with the *L* states (for the details, see *SI Appendix,* Fig. S5). Accordingly, despite the symmetric junction geometry, there arises a stronger (weaker) pileup of electrons or resistivity-dipole formation on the left (right) electrode-molecule interface, which in turn causes more abrupt electrostatic potential drop at the left contact. In view of its importance in semiconductor device applications,[1-3] we can suggest that the nontrivial correlations between the finite-bias voltage drop and the splitting of the QFLs would be a promising target for measurements with atomic-scale spatial resolution. This might be achieved by, *e.g.*, combining the scanning tunneling potentiometry[8-10] and multi-probe scanning tunneling spectroscopy techniques.[24]

## 3. CONCLUSIONS

In summary, we calculated the non-equilibrium QFLs across the nanoscale junctions based on saturated hexanedithiolate (HDT) and conjugated hexatrienedithiolate (HTDT) molecules. The first-principles extraction of QFLs was enabled by the newly developed MS-DFT approach, the key features of which are partitioning the Kohn-Sham states into the left-electrode/channel/right-electrode regions and variationally determining the occupations of channel states. Benchmarking the standard DFT-NEGF calculation results, we first established that during self-consistent non-equilibrium electronic structure calculations it is necessary to adopt an occupation rule that maintains the notion of nonlocal or two separate electrode-originated QFLs within the channel region. For the insulating HDT junction case, we observed the QFLs that are spatially discontinuous at the left and right molecule-electrode interfaces and the accompanying linear electrostatic potential drop and uniformly distributed Landauer residual-resistivity dipoles across the molecular channel. On the other hand, for the conducting HTDT junction counterpart, we observed that the channel-region QFLs split into multiple states originating from the two electrodes and the molecule. Moreover, with the HTDT HOMO entering into the bias window as a good transport channel, we observed the development of a nonlinear electrostatic potential drop and correspondingly the distribution of asymmetric residual-resistivity dipoles in spite of the symmetric junction geometry. Clarifying the nontrivial correlations among the QFL splitting, electrostatic potential drop, and Landauer residual-resistivity dipole formation in non-equilibrium channel regions, our findings underscore the importance of the first-principles extraction of QFLs in nanoscale junctions and point to a new direction for the technology computer-aided design of advanced electronic, optoelectronic, and electrochemical devices.

## 4. MATERIALS AND METHODS

**DFT calculations.** Based on our earlier works,[18, 19, 25] we modeled the HDT- and HTDT-based single-molecule junctions by adopting the *trans* configuration for both molecules and introducing one apex Au atom on top of the 4 × 4 Au(100) slabs. The gap distance between Au electrodes was set to 23.7 Å in terms of the atomic positions of the first Au layers within the *L* and *R* reservoir-region electrodes (red and blue down arrows in Fig. 1C, respectively). We carried out equilibrium DFT calculations using the SIESTA software[26] within the local density approximation.[27] Double ζ-plus-polarization-level numerical atomic orbital basis sets were employed together with the Troullier-Martins type norm-conserving pseudopotentials.[28] The mesh cutoff of 200 Ry for the real-space integration, and 2 × 2 × 1 Monkhorst-Pack *k*-points grid sampling of the Brillouin zone were used. The *C*-region atomic geometries were optimized until the total residual forces on each atom were below 0.02 eV/ Å.

**DFT-NEGF and MS-DFT calculations.** For the finite-bias non-equilibrium electronic structure calculations, we used the DFT-NEGF method implemented within the TranSIESTA code[29] and our in-house MS-DFT method implemented within the SIESTA code.[13] A key feature of MS-DFT that differentiates it from DFT-NEGF is that the self-energy $\Sigma = \tau g_s \tau^\dagger$, where $g_s$ is the surface Green's function and $\tau$ is the electrode–channel coupling matrix, is not introduced within the *self-consistent electronic structure* calculations and postponed until the *post-self-consistent quantum transport* calculations. Remaining within the micro-canonical picture, we invoke the transport-to-optical excitation mapping viewpoint and utilize the space-resolved constrained-search procedure with the constraint of $eV_b = \mu_L - \mu_R$. Then, based on the variational (time-independent) DFT formulations for electronic excitations established by Görling[30] and Levy-Nagy,[31] one can establish a rigorous yet effective DFT formalism for first-principles non-equilibrium electronic structure calculations. We thus complete the finite-bias electronic structure calculations without introducing Σ and explicitly produce the non-equilibrium KS states together with their occupancies.

Once the converged non-equilibrium electronic structure is obtained for a junction using MS-DFT, we finally recover the Landauer or grand-canonical picture as in the DFT-NEGF calculations and calculated its quantum transport properties by invoking the matrix Green's function formalism.[25, 32] The $g_s$ were extracted from separate DFT calculations for four-layer 4 × 4 Au(100) crystals with the 6 × 6 $k_\parallel$-point sampling along the surface $xy$ plane and 10 $k_\perp$ -point sampling along the charge transport $z$ direction. For the comparison of MS-DFT and DFT-NEGF calculation results, we adopted the same $L/C/R$ partitioning scheme (Fig. 1C) and used identical $g_s$. The transmission functions were then obtained according to

$$T(E;V_b) = Tr[\Gamma_L(E;V_b)G(E;V_b)\Gamma_R(E;V_b)G^\dagger(E;V_b)], \quad (7)$$

where $G$ is the retarded Green's function for the channel $C$ and $\Gamma_{L/R} = i(\Sigma_{L/R} - \Sigma_{L/R}^\dagger)$ are the broadening matrices. The numerical convergence of transmission spectra with respect to the $k$-point sampling in the HTDT junction case is provided in *SI Appendix,* Fig. S6. The $I$-$V_b$ characteristics were calculated by invoking the Landauer-Büttiker formula,

$$I(V_b) = \frac{2e}{h} \int_{\mu_L}^{\mu_R} T(E;V_b)[f(E-\mu_R) - f(E-\mu_L)] \, dE. \quad (8)$$

The MPH analyses on DFT-NEGF and MS-DFT calculation output were performed using the Inelastica code.[21]

## Acknowledgments
This work was supported by the Nano-Material Technology Development Program (Nos. 2016M3A7B4024133), Basic Research Program (No. 2017R1A2B3009872), Global Frontier Program (No. 2013M3A6B1078882), and Basic Research Lab Program (No. 2017R1A4A1015400) of the National Research Foundation funded by the Ministry of Science and ICT of Korea. Computational resources were provided by the KISTI Supercomputing Center (KSC-2018-C2-0032).

## Author Contributions
Y.-H.K. developed the theoretical framework and oversaw the project. J.L. implemented the method and J.L. and H.Y. carried out calculations. Y.-H.K., J.L., and H.Y. analyzed the computational results, and Y.-H.K. wrote the manuscript with significant input from J.L.

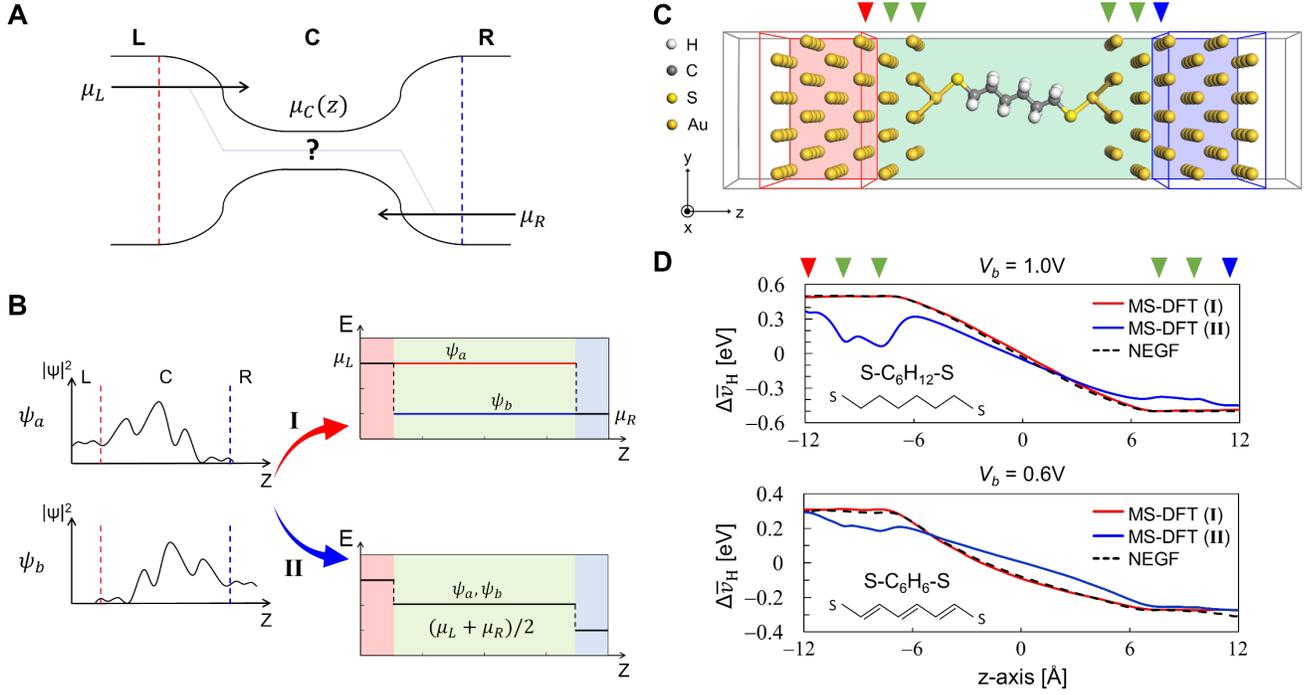

**Fig. 1.** The occupation rule of non-equilibrium channel states within the MS-DFT formalism. (**A**) Schematic of the QFL distribution and charge transport processes in a nanoscale device. According to the Landauer picture, one considers a channel (*C*) sandwiched between the left (*L*) and right (*R*) electrodes. Upon applying a positive voltage $V_b$ to *R*, the electrochemical potentials of *L* and *R* shift to $\mu_L$ and $\mu_R = \mu_L - eV_b$, respectively. The imbalance between right/left-moving electrons (arrows) will drive *C* into a non-equilibrium state and the nature of $\mu_C$ will determine charge transport characteristics. (**B**) Schematics of the occupation rules for the *C* states tested within MS-DFT. Depending on the spatial distribution of each KS wavefunction, we assigned either separate $\mu_L$ and $\mu_R$ (rule **I**) or averaged $(\mu_L + \mu_R)/2$ (rule **II**). (**C**) The optimized structure of the HDT junction model. Red, green, and blue boxes indicate the *L*, *C*, *R* regions, respectively, within the MS-DFT or DFT-NEGF calculations. (**D**) The plane-averaged electrostatic potential difference $\Delta \bar{v}_H$ calculated within DFT-NEGF (black dashed line) and MS-DFT with the occupation rule **I** (red solid lines) or rule **II** (blue solid lines) for the HDT junction at $V_b = 1.0$ V (top panel) and the HTDT junction at $V_b = 0.6$ V (bottom panel). The red, green, and blue down triangles respectively indicate the locations of the *L*-, *C*-, and *R*-region interfacial Au layers marked in (**C**)

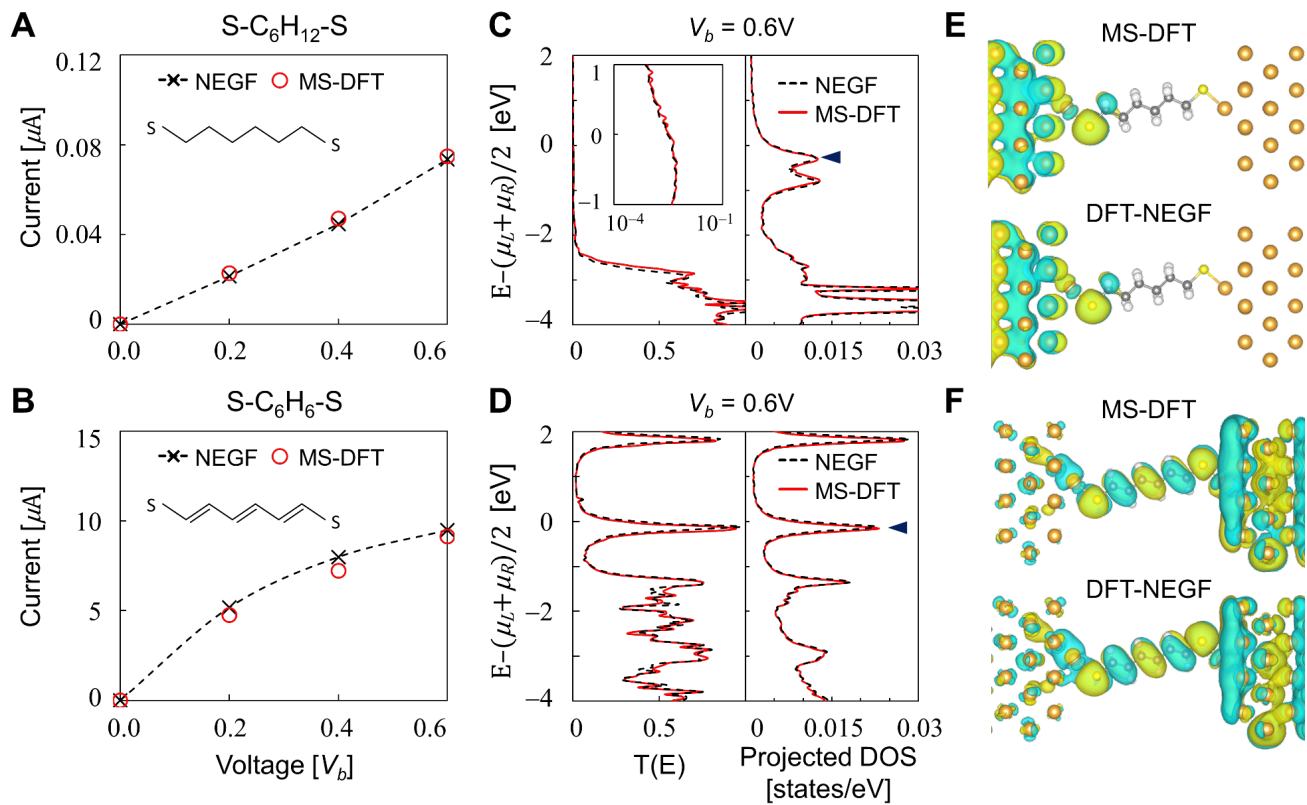

**Fig. 2.** Quantum transport properties of the HDT and HTDT junctions calculated within MS-DFT and DFT-NEGF. Current-bias voltage ($I$-$V_b$) curves of the (**A**) HDT and (**B**) HTDT junctions. Black dashed lines with cross marks and red unfilled circles indicate the results obtained from the DFT-NEGF and MS-DFT calculations, respectively. Insets display the corresponding chemical structures of the molecules. Transmission spectra (left panels) and projected DOS (right panels) of the (**C**) HDT and (**D**) HTDT junctions at $V_b = 0.6$V. Black dashed lines and red solid lines represent the data obtained from the DFT-NEGF and MS-DFT calculations, respectively. In each case, black left triangle indicates the HOMO level. In (**C**), the inset shows the transmission curve in the logarithmic scale. Three-dimensional contour plots of the molecular-projected Hamiltonian HOMO derived from the (**E**) HDT and (**F**) HTDT junctions. In (**E**) and (**F**), the isosurface level is 0.01 Å$^{-3}$.

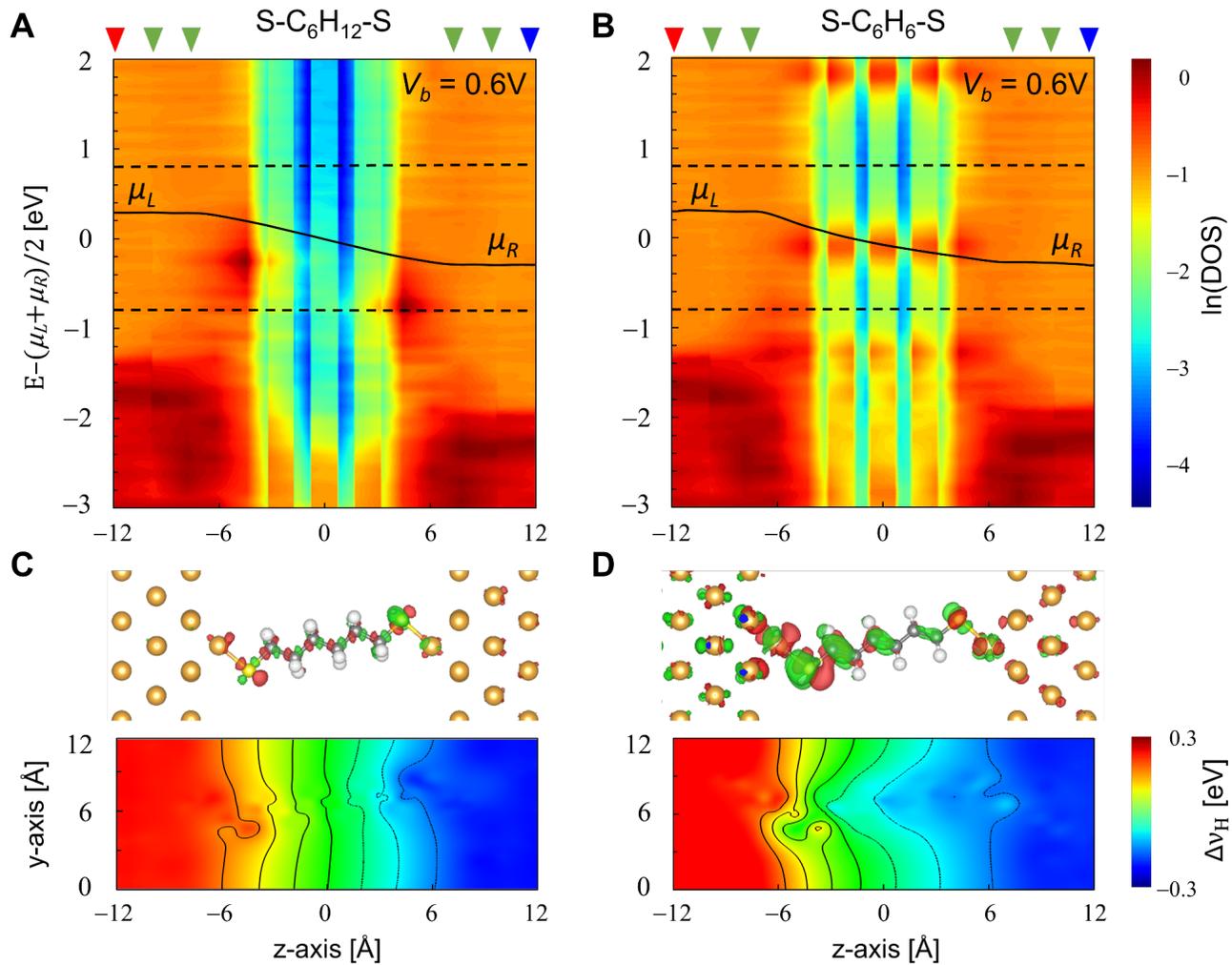

**Fig. 3.** Electronic origin of the single-molecule junctions under a finite bias. The spatially-resolved DOS of the (**A**) HDT and (**B**) HTDT junctions at $V_b = 0.6$ V calculated within MS-DFT. Shown together as black solid lines are plane-averaged electrostatic potential differences $\Delta \bar{v}_H$. The red, green, and blue down triangles respectively indicate the locations of the *L*-, *C*-, and *R*-region interfacial Au layers (Fig. 1C). Three-dimensional contour plots of the Landauer residual-resistivity dipoles overlaid on the atomic structures (top panels) and the two-dimensional contour plots of the corresponding electrostatic potential differences (bottom panels) in the (**C**) HDT and (**D**) HTDT junctions. The isosurface level for the charge density difference plots is 0.0002 e/Å$^{-3}$.

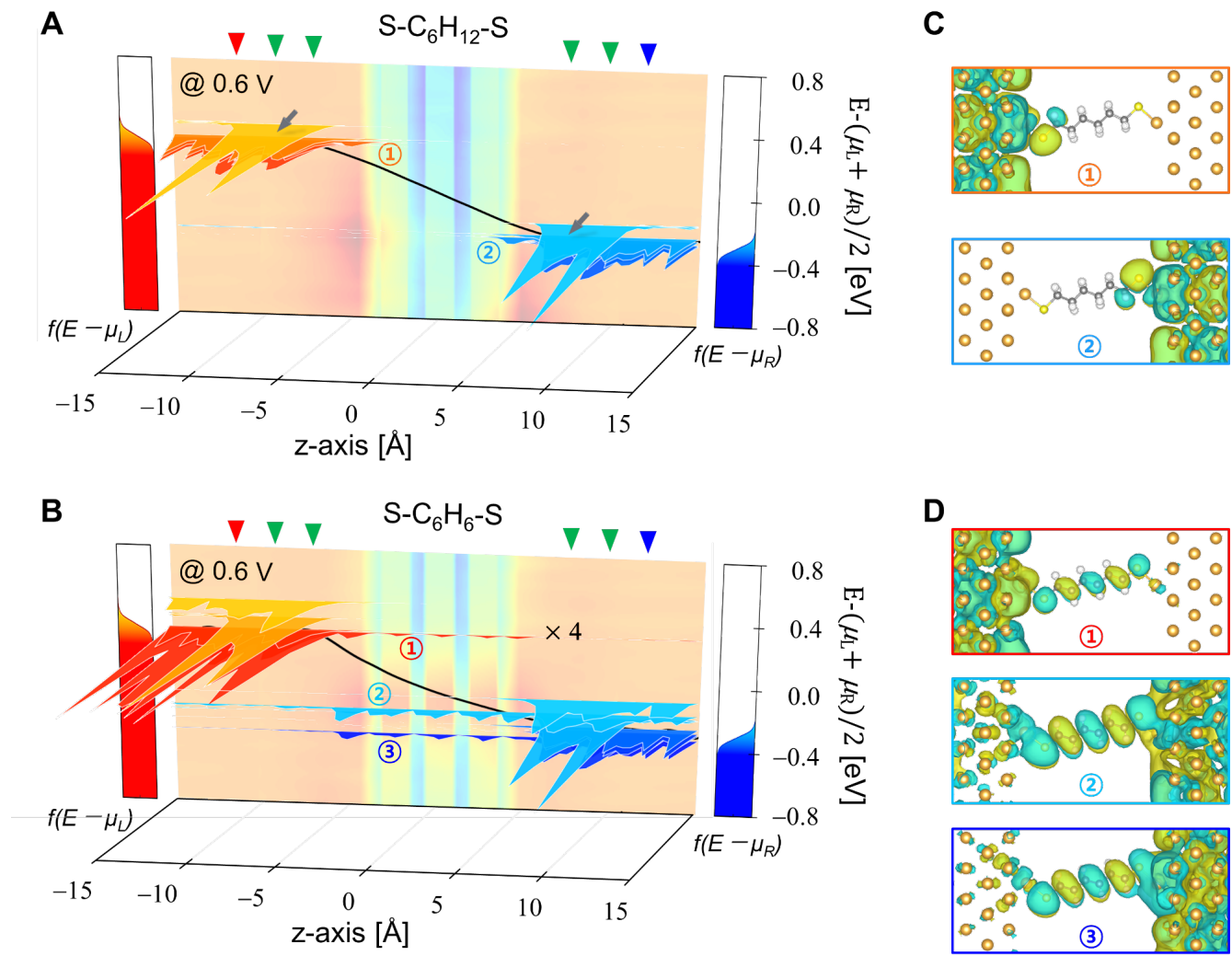

**Fig. 4.** First-principles determination of the QFL profiles across the molecular junctions. The QFL profiles in the (**A**) HDT and (**B**) HTDT junctions at $V_b = 0.6$ V derived from the MS-DFT calculations. The QFL wavefunctions are color-coded according to the electron occupations given by the $f(\varepsilon - \mu_L)$ and $f(\varepsilon - \mu_R)$ for the *L*- and *R*-originated states, respectively. The red, green, and blue down triangles respectively indicate the locations of the *L*-, *C*-, and *R*-region interfacial Au layers (see Fig. 1C). The gray arrows in (**A**) indicate the localized states at metal surfaces. For visual aids, the spatially-resolved DOS shown in Figs. 3A and 3B are included as translucent background images in (**A**) and (**b**), respectively. Three-dimensional contour plots of the most prominent *C*-region QFL wavefunctions (**C**) in the HDT junction (marked as ① and ② in (**A**)) and (**D**) in the HTDT junction (marked as ①, ②, and ③ in (**B**)). In (**C**) and (**D**), the isosurface level is 0.01 Å$^{-3}$.